\documentclass[aps,notitlepage,amsmath,amssymb]{revtex4-1}
\usepackage{graphicx}
\pdfoutput=1

\begin{document}

\title{Probabilistic model of $N$ correlated binary random variables and
non-extensive statistical mechanics}

\author{Julius Ruseckas}

\email{julius.ruseckas@tfai.vu.lt}

\homepage{http://www.itpa.lt/~ruseckas}

\affiliation{Institute of Theoretical Physics and Astronomy, Vilnius University,
A.~Go\v{s}tauto 12, LT-01108 Vilnius, Lithuania}
\begin{abstract}
The framework of non-extensive statistical mechanics, proposed by
Tsallis, has been used to describe a variety of systems. The non-extensive
statistical mechanics is usually introduced in a formal way, thus
simple models exhibiting some important properties described by the
non-extensive statistical mechanics are useful to provide deeper physical
insights. In this article we present a simple model, consisting of
a one-dimensional chain of particles characterized by binary random
variables, that exhibits both the extensivity of the generalized entropy
with $q<1$ and a $q$-Gaussian distribution in the limit of the large
number of particles.
\end{abstract}
\maketitle

\section{introduction}

There exist a number of systems featuring long-range interactions,
long-range memory, and anomalous diffusion, that possess anomalous
properties in view of traditional Boltzmann-Gibbs statistical mechanics.
Non-extensive statistical mechanics is intended to describe such systems
by generalizing the Boltzmann-Gibbs statistics \cite{Tsallis2009-1,Tsallis2009-2,Telesca2010}.
In general, the non-extensive statistical mechanics can be applied
to describe the systems that, depending on the initial conditions,
are not ergodic in the entire phase space and may prefer a particular
subspace which has a scale invariant geometry, a hierarchical or multifractal
structure. Concepts related to the non-extensive statistical mechanics
have found applications in a variety of disciplines: physics, chemistry,
biology, mathematics, economics, and informatics \cite{Gell-Mann2004,Abe2006,Picoli2009}.

The non-extensive statistical mechanics is based on a generalized
entropy \cite{Tsallis2009-1} 
\begin{equation}
S_{q}=\frac{1-\int[p(x)]^{q}dx}{q-1}\,,\label{eq:q-entr}
\end{equation}
where $p(x)$ is a probability density function of finding the system
in the state characterized by the parameter $x$, while $q$ is a
parameter describing the non-extensiveness of the system. Entropy
(\ref{eq:q-entr}) is an extension of the Boltzmann-Gibbs entropy
\begin{equation}
S_{\mathrm{BG}}=-\int p(x)\ln p(x)dx\label{eq:BG}
\end{equation}
which is recovered from Eq.~(\ref{eq:q-entr}) in the limit $q\rightarrow1$
\cite{Tsallis2009-1,Tsallis2009-2}. More generalized entropies and
distribution functions are introduced in Refs.~\cite{Hanel2011-1,Hanel2011-2}.
Statistics associated to Eq.~(\ref{eq:q-entr}) has been successfully
applied to phenomena with the scale-invariant geometry, like in low-dimensional
dissipative and conservative maps in the dynamical systems \cite{Afsar2013,Tirnakli2009,Ruiz2012},
anomalous diffusion \cite{Huang2010,Prehl2012}, turbulent flows \cite{Beck2013},
Langevin dynamics with fluctuating temperature \cite{Budini2012,Du2012},
spin-glasses \cite{Pickup2009}, plasma \cite{Liu2008} and to the
financial systems \cite{Borland2012,Drozdz2010,Gontis2010}.

By maximizing the entropy (\ref{eq:q-entr}) with the constraints
$\int_{-\infty}^{+\infty}p(x)dx=1$ and
\begin{equation}
\frac{\int_{-\infty}^{+\infty}x^{2}[p(x)]^{q}dx}{\int_{-\infty}^{+\infty}[p(x)]^{q}dx}=\sigma_{q}^{2}\,,
\end{equation}
where $\sigma_{q}^{2}$ is the generalized second-order moment \cite{Tsallis1998,Prato1999,Tsallis1999},
one obtains the $q$-Gaussian distribution density
\begin{equation}
p_{q}(x)=C\exp_{q}(-A_{q}x^{2})\,.\label{eq:q-gauss}
\end{equation}
Here $\exp_{q}(\cdot)$ is the $q$-exponential function, defined
as 
\begin{equation}
\exp_{q}(x)\equiv[1+(1-q)x]_{+}^{\frac{1}{1-q}}\,,\label{eq:q-exp1}
\end{equation}
with $[x]_{+}=x$ if $x>0$, and $[x]_{+}=0$ otherwise. The $q$-Gaussian
distribution (or distribution very close to it) appears in many physical
systems, such as cold atoms in dissipative optical lattices \cite{Douglas2006},
dusty plasma \cite{Liu2008}, motion of hydra cells \cite{Upaddhyaya2001},
and defect turbulence \cite{Daniels2004}. The $q$-Gaussian distribution
is one of the most important distributions in the non-extensive statistical
mechanics. It's importance stems from the generalized central limit
theorems \cite{Umarov2008,Umarov2010,Umarov2007a}. According to $q$-generalized
central limit theorem, $q$-Gaussian can result from a sum of $N$
$q$-independent random variables. The $q$-independence is defined
in \cite{Umarov2008} through the $q$-product \cite{Nivanen2003,Borges2004},
and the $q$-generalized Fourier transform \cite{Umarov2008}. When
$q\neq1$, $q$-independence corresponds to a global correlation of
the $N$ random variables. However, the rigorous definition of $q$-independence
is not transparent enough in physical terms.

The non-extensive statistical mechanics is introduced in a formal
way, starting from the maximization of the generalized entropy \cite{Tsallis2009-1}.
Therefore, simple models providing some degree of intuition about
non-extensive statistical mechanics can be useful for understanding
it. There has been some effort to create such simple models. In Ref.~\cite{Tsallis2005}
a system composed of $N$ distinguishable particles, each particle
characterized by a binary random variable, has been constructed so
that the number of states with non-zero probability grows with the
number of particles $N$ not exponentially, but as a power law. For
such a system in the limit $N\rightarrow\infty$ the ratio $S_{q}(N)/N$
is finite not for the Boltzmann-Gibbs entropy but for the generalized
entropy with some specific value of $q$. The starting point in the
construction is the Leibnitz triangle, then initial probabilities
are redistributed into a small number of all the other possible states,
in such a way that the norm is preserved. For example, in the restricted
uniform model \cite{Tsallis2005} for a fixed value of $N$ all nonvanishing
probabilities are equal. In the proposed models that yield $q\neq1$
there are $d+1$ no-zero probabilities and the value of $q$ is given
by $q=1-1/d$.

In Refs.~\cite{Moyano2006,Rodriguez2008}, the goal has been to construct
simple models providing $q$-Gaussian distributions. As in \cite{Tsallis2005},
the models considered in \cite{Moyano2006} consist of $N$ independent
and distinguishable binary variables, each of them having two equally
probable states. The models presented in \cite{Moyano2006} are strictly
scale-invariant, however, they do not approach a $q$-Gaussian form
when the number of particles $N$ in the model increases \cite{Hilhorst2007}.
The situation is different with the models presented in \cite{Rodriguez2008}:
the two proposed models do approach a $q$-Gaussian form, the second
of them does so by construction. All models in \cite{Moyano2006,Rodriguez2008},
except the last model of \cite{Rodriguez2008} are for $q\leq1$.
The drawback of the models from Ref.~\cite{Rodriguez2008} is that
the standard Boltzmann-Gibbs entropy remains extensive. In addition,
the models are constructed artificially and it is hard to see how
they can be related to real physical systems. 

The goal of this paper is to provide a simple model that achieves
both the extensivity of the generalized entropy with $q\neq1$ and
$q$-Gaussian distribution in the limit of the large number of particles.
In addition, we want to construct a model that is closer to situations
in physical systems. We expect that such a model can provide deeper
insights into non-extensive statistical mechanics than the previously
constructed simple models.

The paper is organized as follows: To highlight differences from our
proposed model, a simple model consisting of uncorrelated binary random
variables and leading to extensive Boltzmann-Gibbs entropy and a Gaussian
distribution is presented in Section~\ref{sec:uncorrelated}. In
Section~\ref{sec:correlated} we construct a simple model exhibiting
the extensivity of the generalized entropy with $q\neq1$ and $q$-Gaussian
distribution in the limit of the large number of particles. Section~\ref{sec:conclusions}
summarizes our findings.

\section{Model of uncorrelated binary random variables}

\label{sec:uncorrelated}At first let us consider a model consisting
from $N$ uncorrelated binary random variables. Physical implementation
of such a model could be $N$ particles of spin $\frac{1}{2}$, the
projection of each spin to the $z$ axis can acquire the values $\pm\frac{1}{2}$.
The microscopic configuration of the system can be described by a
sequence of spin projections $s_{1}s_{2}\ldots s_{N}$, where each
$s_{i}=\pm\frac{1}{2}$. There are $W=2^{N}$ different microscopic
configurations. As is usual in statistical mechanics for the description
of a microcanonical ensemble, we assign to each microscopic configuration
the same probability. Thus the probability of each microscopic configuration
is 
\begin{equation}
P=\frac{1}{W}=\frac{1}{2^{N}}\,.\label{eq:equiprob}
\end{equation}
Note, that this system has a property of composability: if we have
two spin chains with $W_{1}$ and $W_{2}$ microscopic configurations,
then we can join them to form a larger system. The description of
a larger system is just concatenation of the descriptions of each
subsystems and the number of microscopic configurations of the whole
system is $W=W_{1}W_{2}$. The standard Boltzmann-Gibbs entropy $S_{\mathrm{BG}}=k_{\mathrm{B}}\ln W$
is extensive for this system: $S_{\mathrm{BG}}=Nk_{\mathrm{B}}\ln2$
grows linearly with $N$.

Let us consider a macroscopic quantity, the total spin of the system
\begin{equation}
M=\sum_{i=1}^{N}s_{i}\,.
\end{equation}
The total spin can take values $M=-\frac{N}{2},-\frac{N}{2}+1,\ldots,\frac{N}{2}-1,\frac{N}{2}$.
The value of $M$ can be obtained when there are $n=M+\frac{N}{2}$
spins with the projection $+\frac{1}{2}$, the remaining spins have
projection $-\frac{1}{2}$. The macroscopic configuration corresponding
to the given value of $M$ can be realized by $\frac{N!}{n!(N-n)!}$
microscopic configurations, thus using Eq.~(\ref{eq:equiprob}) the
probability of each macroscopic configuration is
\begin{equation}
P_{M}=\frac{1}{2^{N}}\frac{N!}{n!(N-n)!}\,.\label{eq:prob-macro}
\end{equation}
Note, that the probabilities of macroscopic configurations are normalized:
\begin{equation}
\sum_{M=-N}^{N}P_{M}=1\,.
\end{equation}
Using Eq.~(\ref{eq:prob-macro}) we can calculate the average spin
of the system $\langle M\rangle=0$ and the standard deviation 
\begin{equation}
\sqrt{\langle M^{2}\rangle-\langle M\rangle^{2}}=\frac{\sqrt{N}}{2}\,.\label{eq:deviation}
\end{equation}
From Eq.~(\ref{eq:deviation}) it follows that the relative width
of the distribution of the total spin $M$ decreases with number of
spins $N$ as $\frac{1}{\sqrt{N}}$. When $N$ is large then we can
approximate the factorials using Stirling formula
\begin{equation}
n!\approx\sqrt{2\pi n}n^{n}e^{-n}
\end{equation}
and obtain a Gaussian distribution
\begin{equation}
P_{M}\approx\frac{1}{\sqrt{\pi\frac{N}{2}}}e^{-\frac{2M^{2}}{N}}
\end{equation}
The Gaussian distribution can be obtained by maximizing the Boltzmann-Gibbs
entropy (\ref{eq:BG}) with appropriate constraints.

\section{Model of correlated spins}

\begin{figure}

\includegraphics[width=0.4\textwidth]{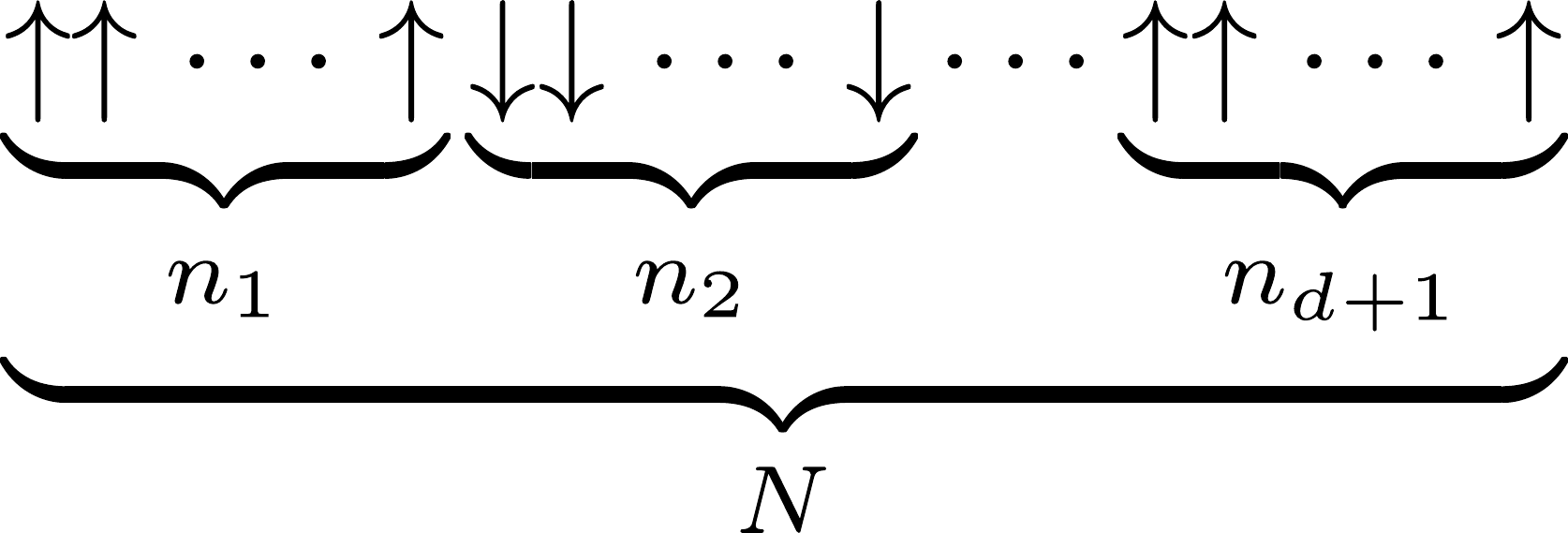}\protect\caption{One-dimensional spin chain having $N$ spins. There are $d$ spin
flips, so that the spin chain consists of $d+1$ domains of spins
pointing to the same direction. The lengths of the domains are $n_{1},n_{2},\ldots,n_{d+1}$.}
\label{fig:spin-chain}
\end{figure}

\label{sec:correlated}In this Section we investigate a system consisting
from $N$ correlated binary random variables. Similarly as in the
previous Section, we can think about a one-dimensional spin chain
consisting from $N$ spins. However, the spins are correlated: spins
next to each other have almost always the same direction, except there
are $d$ cases when the next spin has an opposite direction, as is
shown in Fig.~\ref{fig:spin-chain}. Thus the spin chain consists
from $d+1$ domains with spins pointing in the same direction and
has $d$ boundaries between domains. This model is inspired by a connection
between non-extensive statistics and critical phenomena \cite{Robledo1999,Robledo2005,Robledo2007}.
As it has been shown in Refs.~\cite{Robledo1999,Robledo2005,Robledo2007},
the properties of a single large cluster of the order parameter at
a critical point in thermal systems can be described by non-extensive
statistics. The restriction of the number of allowed states in our
model is very similar to the models presented in \cite{Tsallis2005}.

Let us calculate the number of allowed microscopic configurations
of the spin chain. If the length of $i$-th domain is $n_{i}$ then
we need to calculate the number of possible partitions such that
\begin{equation}
\sum_{i=1}^{d+1}n_{i}=N\,.
\end{equation}
This number is equivalent to the number of ways one can place $d$
domain boundaries into $N-1$ possible positions. Since the spins
in the first domain can be up or down, the number of microscopic configurations
is twice as large. Thus the number of allowed microscopic configurations
of the spin chain is
\begin{equation}
W=\frac{2(N-1)!}{d!(N-d-1)!}=\frac{2}{d!}(N-1)\cdots(N-d)\,.\label{eq:w-exact}
\end{equation}
For large $N\gg d$ we have that the number of microscopic configurations
grows as a power-law of the number of spins, not exponentially: 
\begin{equation}
W\sim\frac{2}{d!}N^{d}\,.\label{eq:w-macro}
\end{equation}
Assigning to each allowed microscopic configuration the same probability
$P=1/W$ we get that in this situation the traditional Boltzmann-Gibbs
entropy is not linearly proportional to $N$ and thus is not extensive.
The generalized entropy Eq.~(\ref{eq:q-entr}) for equal probabilities
$1/W$ takes the form
\begin{equation}
S_{q}=k_{\mathrm{B}}\frac{1-W^{1-q}}{q-1}\,.\label{eq:q-w}
\end{equation}
Using Eqs.~(\ref{eq:w-macro}) and (\ref{eq:q-w}) on gets that the
generalized entropy is extensive (proportional to $N$) only when
$q$ is
\begin{equation}
q_{\mathrm{stat}}=1-\frac{1}{d}\,.\label{eq:q-stat}
\end{equation}
This is the same dependency of $q$ on $d$ as in \cite{Tsallis2005}.

\begin{figure}
\includegraphics[width=0.6\textwidth]{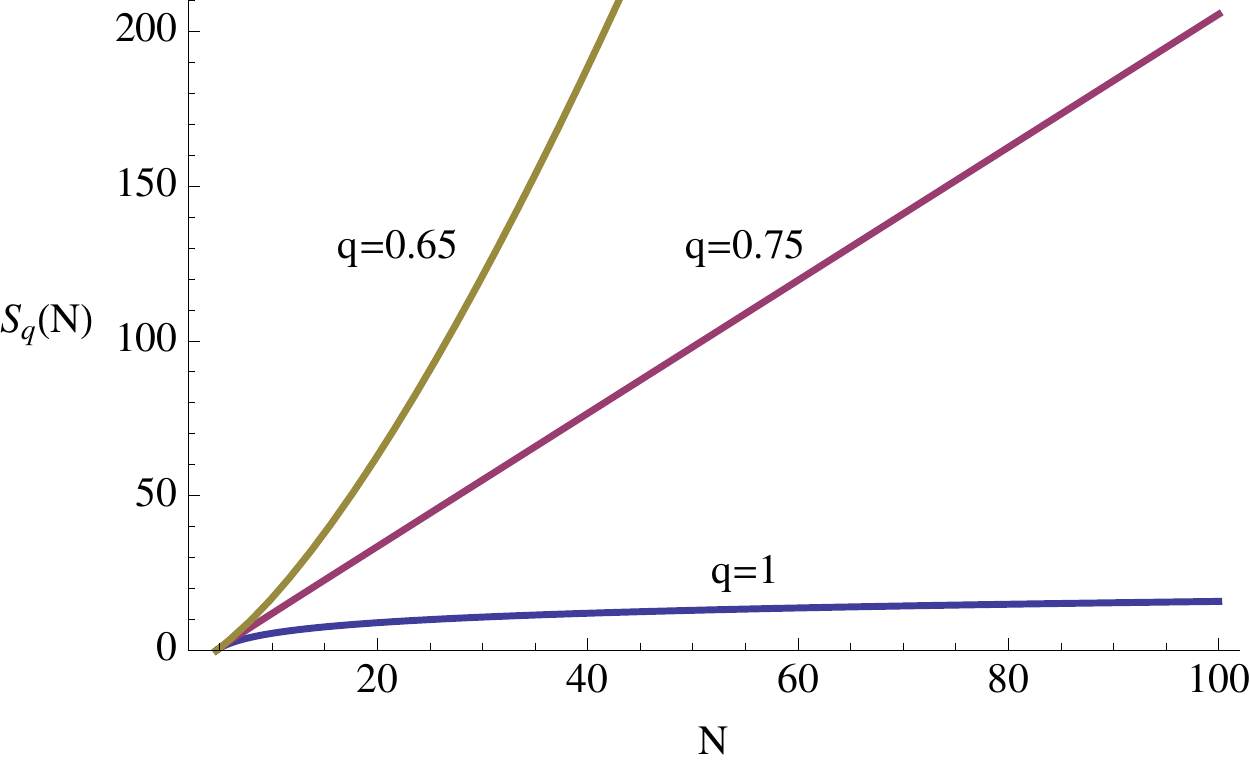}\protect\caption{Dependence of the generalized entropy $S_{q}$ on the number of particles
$N$ in the spin chain for various values of $q$. The number of domain
walls $d=4$. Only for $q=q_{\mathrm{stat}}=1-1/d=0.75$ the limit
$\lim_{N\rightarrow\infty}S_{q}(N)/N$ has a finite value; the limit
vanishes (diverges) for $q>0.75$ ($q<0.75)$.}
\label{fig:sq-n}

\end{figure}

In Fig.~\ref{fig:sq-n} the dependence of the generalized entropy
(\ref{eq:q-w}) on the size $N$ of the spin chain for various values
of $q$ is shown. As one can see, only for $q=q_{\mathrm{stat}}$
the limit $\lim_{N\rightarrow\infty}S_{q}/N$ is finite, this limit
vanishes or diverges for all other values of $q$. Thus our simple
model can be characterized by the generalized entropy with $q<1$.

Now let us consider the macroscopic variable, the total spin $M$.
If there are $d+1$ domains and the total spin is $M$ then we have
the equality
\begin{equation}
s_{1}\sum_{i=1}^{d+1}(-1)^{i-1}n_{i}=M\,.\label{eq:m-1}
\end{equation}
Eq.~(\ref{eq:m-1}) can be written as
\begin{equation}
N_{\mathrm{odd}}-N_{\mathrm{even}}=\frac{M}{s_{1}}\,,\label{eq:o-e-1}
\end{equation}
where 
\begin{equation}
N_{\mathrm{odd}}=\sum_{k=0}^{\left\lfloor \frac{d}{2}\right\rfloor }n_{1+2k}
\end{equation}
is the total length of odd-numbered domains, 
\begin{equation}
N_{\mathrm{even}}=\sum_{k=1}^{\left\lfloor \frac{d+1}{2}\right\rfloor }n_{2k}
\end{equation}
is the total length of even-numbered domains. Here $\left\lfloor \cdot\right\rfloor $
denotes an integer part of a number. In addition, the total lengths
of odd- and even-numbered domains should obey the equation 
\begin{equation}
N_{\mathrm{odd}}+N_{\mathrm{even}}=N\,.\label{eq:o-e-2}
\end{equation}
Eqs.~(\ref{eq:o-e-1}) and (\ref{eq:o-e-2}) have only one solution
\begin{equation}
N_{\mathrm{odd}}=\frac{1}{2}\left(N+\frac{M}{s_{1}}\right)\,,\qquad N_{\mathrm{even}}=\frac{1}{2}\left(N-\frac{M}{s_{1}}\right)\,.\label{eq:n-odd-even}
\end{equation}
The total length of odd-numbered domains $N_{\mathrm{odd}}$ is a
sum of $\left\lfloor \frac{d}{2}\right\rfloor +1$ terms, the total
length of even-numbered domains $N_{\mathrm{even}}$ is a sum of $\left\lfloor \frac{d+1}{2}\right\rfloor $
terms. Similarly as in Eq.~(\ref{eq:w-exact}), the number of ways
to choose $n_{1},n_{2},\ldots,n_{d+1}$ is equal to the number of
ways to place $\left\lfloor \frac{d}{2}\right\rfloor $ domain boundaries
into $N_{\mathrm{odd}}-1$ positions multiplied by the number of ways
to place $\left\lfloor \frac{d+1}{2}\right\rfloor -1$ domain boundaries
into $N_{\mathrm{even}}-1$ positions. Thus the number of ways to
choose the domain lengths when $M$, $N$ and $s_{1}$ are given is
\begin{equation}
W(N,M,s_{1})=\frac{1}{\left(\left\lfloor \frac{d}{2}\right\rfloor \right)!}\biggl(N_{\mathrm{odd}}-1\biggr)\cdots\biggl(N_{\mathrm{odd}}-\left\lfloor \frac{d}{2}\right\rfloor \biggr)\frac{1}{\left(\left\lfloor \frac{d-1}{2}\right\rfloor \right)!}\biggl(N_{\mathrm{even}}-1\biggr)\cdots\biggl(N_{\mathrm{even}}-\left\lfloor \frac{d-1}{2}\right\rfloor \biggr)\,.\label{eq:w-n-s1}
\end{equation}
The number of microscopic configurations corresponding to a given
value of $M$ is
\begin{equation}
W(N,M)=\sum_{s_{1}=\pm1/2}W(N,M,s_{1})\,.\label{eq:w-n-m}
\end{equation}
Using Eqs.~(\ref{eq:n-odd-even})--(\ref{eq:w-n-m}) we obtain that
the number of microscopic configurations corresponding to a given
value of $M$ is 
\begin{equation}
W(N,M)=\frac{1}{2^{d-2}\left[\left(\frac{d-1}{2}\right)!\right]^{2}}\left((N-2)^{2}-4M^{2}\right)\cdots\left((N-(d-1))^{2}-4M^{2}\right)\label{eq:w-odd}
\end{equation}
when $d$ is odd and
\begin{equation}
W(N,M)=\frac{(N-d)}{2^{d-2}\frac{d}{2}\left[\left(\frac{d}{2}-1\right)!\right]^{2}}\left((N-2)^{2}-4M^{2}\right)\cdots\left((N-(d-2))^{2}-4M^{2}\right)\label{eq:w-even}
\end{equation}
when $d$ is even. When $N\gg d$ then Eqs.~(\ref{eq:w-odd}) and
(\ref{eq:w-even}) can be approximated as
\begin{equation}
W(N,M)\approx\frac{N^{d-1}}{2^{d-2}\left(\left\lfloor \frac{d}{2}\right\rfloor \right)!\left(\left\lfloor \frac{d-1}{2}\right\rfloor \right)!}\left(1-\frac{4M^{2}}{N^{2}}\right)^{\left\lfloor \frac{d-1}{2}\right\rfloor }\,.\label{eq:w-approx}
\end{equation}
Instead of the total spin $M$ it is convenient to consider a scaled
variable
\begin{equation}
x=\frac{2M}{N}\,.
\end{equation}
From Eq.~(\ref{eq:w-approx}) we get that the distribution $P_{x}(x)$
of the variable $x$ for large $N$ is proportional to 
\begin{equation}
P_{x}(x)\propto(1-x^{2})^{\left\lfloor \frac{d-1}{2}\right\rfloor }\,.\label{eq:px}
\end{equation}

When $q<1$ then the $q$-Gaussian distribution (\ref{eq:q-gauss})
has compact support, the range of possible values of $x$ is limited
by the condition $|x|\leqslant x_{q}$, where
\begin{equation}
x_{q}=\frac{1}{\sqrt{(1-q)A_{q}}}\,.
\end{equation}
Using the limiting value $x_{q}$ the expression (\ref{eq:q-gauss})
for the $q$-Gaussian distribution takes the form
\begin{equation}
p_{q}(x)=\frac{\Gamma\left(\frac{5-3q}{2(1-q)}\right)}{\sqrt{\pi}x_{q}\Gamma\left(\frac{2-q}{1-q}\right)}\left[1-\frac{x^{2}}{x_{q}^{2}}\right]_{+}^{\frac{1}{1-q}}\,.\label{eq:q-gauss-xq}
\end{equation}
By rescaling the variable $x$ the expression (\ref{eq:q-gauss-xq})
for $q$-Gaussian distribution in the case of $q<1$ can be written
as
\begin{equation}
p_{q}(x)\propto(1-x^{2})^{\frac{1}{1-q}}\,.\label{eq:q-gauss-2}
\end{equation}
Comparing Eq.~(\ref{eq:px}) with Eq.~(\ref{eq:q-gauss-2}) we see
that the distribution of the total spin $M$ in our model in the limit
of large number of spins $N$ is a $q$-Gaussian with
\begin{equation}
q_{\mathrm{dist}}=1-\frac{1}{\left\lfloor \frac{d-1}{2}\right\rfloor }\,.\label{eq:q-dist}
\end{equation}

\begin{figure}
\includegraphics[width=0.6\textwidth]{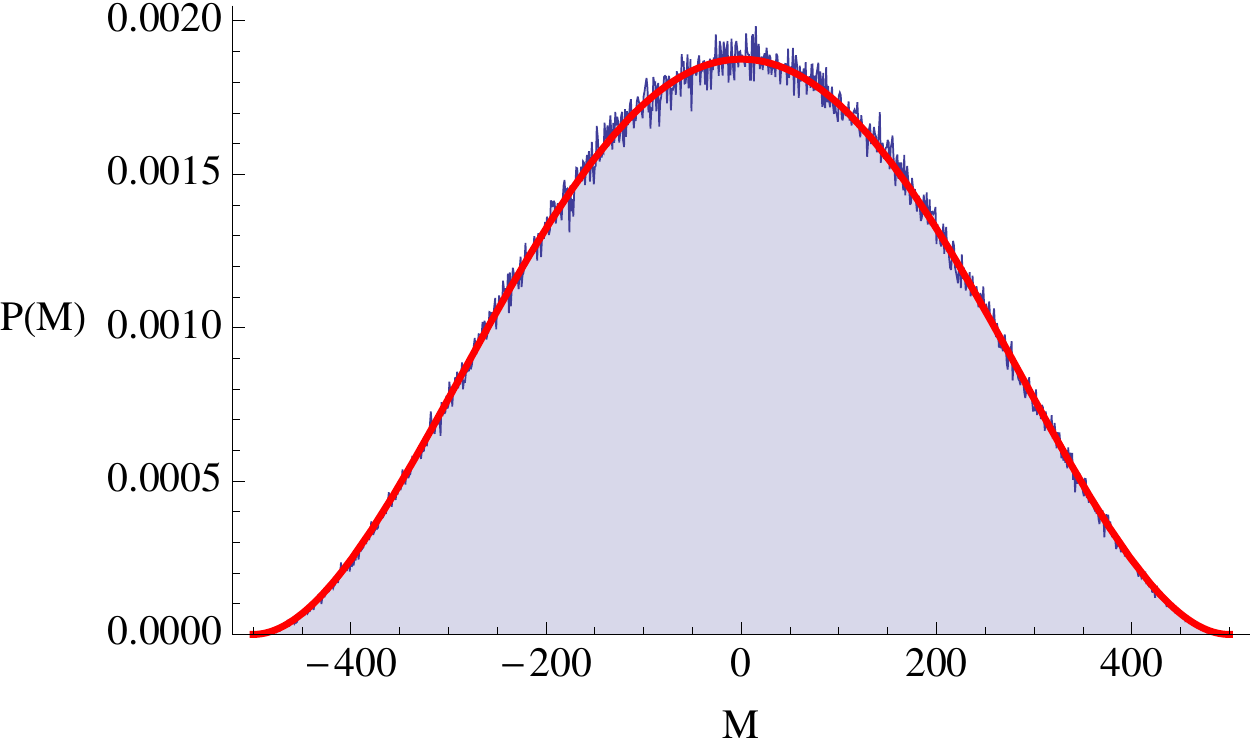}\protect\caption{Comparison of numerically obtained histogram of total spin $M$ (gray
area) with $q$-Gaussian distribution (solid red line). The histogram
is calculated using an ensemble of randomly generated spin chains
that have the length $N=1000$ spins and $d=5$ spin flips. The $q$-Gaussian
distribution is given by Eq.~(\ref{eq:q-gauss-xq}) with $x_{q}=N/2$.}
\label{fig:distr}
\end{figure}

Comparison of the histogram of the total spin $M$, calculated using
an ans amble of randomly generated spin chains, with a $q$-Gaussian
distribution (\ref{eq:q-gauss-xq}) is shown in Fig.~\ref{fig:distr}.
We see that the $q$-Gaussian describes the distribution of the total
spin very well. By increasing the number of spin flips $d$, both
$q_{\mathrm{stat}}$ and $q_{\mathrm{dist}}$ approaches the value
of $1$, as follows from Eqs.~(\ref{eq:q-stat}) and (\ref{eq:q-dist}).
In the limit of $d\rightarrow\infty$ the distribution of total spin
becomes Gaussian.

From Eqs.~(\ref{eq:q-stat}) and (\ref{eq:q-dist}) follows the relationship
between the two $q$ values:
\begin{equation}
\frac{1}{1-q_{\mathrm{dist}}}=\left\lfloor \frac{1}{2}\left(\frac{1}{1-q_{\mathrm{stat}}}-1\right)\right\rfloor \,.
\end{equation}
This equation is similar to relations presented in Eq.~(18) of Ref.~\cite{Ruiz2012a},
in the footnote in page 15378 of Ref.~\cite{Tsallis2005}, and Eq.
(5) in Ref.~\cite{Celikoglu2010}.

\section{Discussion}

\label{sec:conclusions}Our simple model of a spin chain exhibits
both non-extensive behavior of Boltzmann-Gibbs entropy and $q$-Gaussian
distribution of the total spin. This is in contrast to the other models
involving $N$ binary random variables: the models presented in \cite{Tsallis2005}
demonstrate that the generalized entropy with $q\neq1$ can be extensive,
but they do not provide $q$-Gaussian distribution, whereas the models
from Ref.~\cite{Rodriguez2008} yield $q$-Gaussians but for them
the Boltzmann-Gibbs entropy is extensive. Thus the model, presented
in this paper, is an improvement over earlier models and may provide
deeper insights into non-extensive statistical mechanics.

By comparing the model with the chain of uncorrelated spins, presented
in Section \ref{sec:uncorrelated}, we see that the reason for the
non-extensivity of Boltzmann-Gibbs entropy and the extensivity of
the generalized entropy is the reduction of the number of allowed
microscopic configurations, resulting from the restriction of the
number of allowed spin flips. Similar reason is behind the differences
in the distribution of the total spin $M$: the number of possible
configurations having $N_{\mathrm{odd}}$ spins pointing in one direction
and $N_{\mathrm{even}}$ spins in the opposite direction in the chain
of uncorrelated spins is the exponential function of $N_{\mathrm{odd}}N_{\mathrm{even}}$,
resulting in the Gaussian function of $M$. When the number of spin
flips is restricted, the number of configurations is a power-law function
of $N_{\mathrm{odd}}N_{\mathrm{even}}$, resulting in the $q$-Gaussian.

The model can be slightly modified by requiring that the number of
spin flips is not constant, but can be a random number not larger
than $d$. Since the number of microscopic configurations having $d$
spin flips grows as $N^{d}$, the contribution of configurations with
smaller number of spin flips becomes negligible in the limit of large
$N$. Thus the conclusions of Section~\ref{sec:correlated} remain
valid also for this modified model.

Note, that the values of $q$ obtained from the entropy (\ref{eq:q-stat})
and from the distribution (\ref{eq:q-dist}) are different. This difference
is not surprising and is similar to the $q$-triplet in the non-extensive
statistical mechanics. Usually the systems described by the non-extensive
statistical mechanics have three different values of $q$, the $q$-triplet
\cite{Tsallis2009-1}. This triplet consist from the values of $q$,
($q_{\mathrm{sen}}$, $q_{\mathrm{rel}}$,$q_{\mathrm{stat}}$) obtained
from the sensitivity to the initial conditions, the relaxation in
phase-space, and the distribution of energies at a stationary state
\cite{Tsallis2009-1}. Since our model does not involve the energy
and there is no evolution in time, we have obtained only two values
$q$. It would be interesting task for the future to extend the model
and get the full $q$-triplet.

The model presented here provides $q$-Gaussian distribution with
$q<1$. Thus another open question is whether it is possible to modify
the model to obtain $q$-Gaussian distribution with $q>1$.

\end{document}